\documentclass{article}
\usepackage[T1]{fontenc} 
\usepackage[utf8]{inputenc} 
\usepackage{ismir,amsmath,cite,url}
\usepackage{graphicx}
\usepackage{color}
\usepackage{xcolor}
\definecolor{RYB1}{RGB}{218,232,252}
\usepackage{array}
\newcommand{\PreserveBackslash}[1]{\let\temp=\\#1\let\\=\temp}
\newcolumntype{C}[1]{>{\PreserveBackslash\centering}p{#1}}
\newcolumntype{R}[1]{>{\PreserveBackslash\raggedleft}p{#1}}
\newcolumntype{L}[1]{>{\PreserveBackslash\raggedright}p{#1}}
\usepackage{multirow}
\usepackage{pgfplots}

\usepackage{lineno}

\title{Singing beat tracking with Self-supervised front-end and linear transformers}

\threeauthors
  {Mojtaba Heydari} {XXXXXXXXXXXXXXXXXXXXXX \\ {\tt mheydari@ur.rochester.edu}}
  {     } {\bf      }
  {Zhiyao Duan} {XXXXXXXXXXXXXXXXXXXXXX \\ {\tt  zhiyao.duan@rochester.edu}}

\oneauthor
   {\hspace{-1cm} Mojtaba Heydari \hspace{4cm}     Zhiyao Duan} {Department of Electrical and Computer Engineering \\
University of Rochester, 500 Wilson Blvd, Rochester, NY 14627, USA \\ {\tt \( \)mheydari\( \)@ur.rochester.edu \text{} \hspace{0.2cm}  zhiyao.duan@rochester.edu}}

\def\authorname{M. Heydari and Z. Duan}

\begin{document}

\maketitle
\begin{abstract}
Tracking beats of singing voices without the presence of musical accompaniment can find many applications in music production, automatic song arrangement, and social media interaction.
Its main challenge is the lack of strong rhythmic and harmonic patterns that are important for music rhythmic analysis in general. Even for human listeners, this can be a challenging task. As a result, existing music beat tracking systems fail to deliver satisfactory performance on singing voices. In this paper, we propose singing beat tracking as a novel task, and propose the first approach to solving this task. Our approach leverages semantic information of singing voices by employing pre-trained self-supervised WavLM and DistilHuBERT speech representations as the front-end and uses a self-attention encoder layer to predict beats. To train and test the system, we obtain separated singing voices and their beat annotations using source separation and beat tracking on complete songs, followed by manual corrections. 
Experiments on the 741 separated vocal tracks of the GTZAN dataset show that the proposed system outperforms several state-of-the-art music beat tracking methods by a large margin in terms of beat tracking accuracy. Ablation studies 
also confirm the advantages of pre-trained self-supervised speech representations over generic spectral features.

\end{abstract}
\section{Introduction}\label{sec:introduction}

Music tempo and beat detection are two of the core and well-defined MIR topics, and scholars have proposed many approaches to addressing different aspects of them for various music genres. For instance, some works such as ~\cite{ellis2007beat,dixon2007evaluation,davies2009evaluation,mottaghi2017obtain} proposed some unsupervised approaches to detect music beat and tempo by using some low-level features like onset strengths. More recently, several more recent approaches employ neural networks to address the mentioned tasks~\cite{bock2020deconstruct,heydari2021beatnet,Bock:1,Heydari}. Extracting singing rhythmic parameters makes it possible to address several MIR problems such as automatic instrumental and singing tracks alignment, automatic music mix, and remix, interactive content creation on social media platforms, etc.

Many of the proposed music rhythmic analysis approaches~\cite{ellis2007beat,bock2020deconstruct,Bock:1} perform in an offline fashion while others~\cite{heydari2021beatnet,oliveira2010ibt,heydari2022novel} are capable of extracting music rhythmic features causally and even in real-time.


In contrast to all of the mentioned models, beat and tempo detection for singing voice is an untapped MIR task.  There are significant inherent differences between the natures of complete music and singing voices. One of those differences is that complete music pieces usually contain rich percussive and harmonic profiles while singing tracks usually lack such beneficial parameters making their rhythmic analysis more demanding compared to the former group. Their other important difference is that music beat tracking models usually only use acoustical clues such as magnitude spectrogram as input features while singing tracks are more similar to speech signals where the models may require to deal with para-linguistics, semantic, and phonemic level inputs in addition to acoustical features.

Our contributions in this paper are as follows:

1- We introduce singing beat tracking as a novel MIR task. We propose two strategies to tackle the lack of annotated data for this task by leveraging pre-existing datasets and beat tracking and source separation techniques. We also propose a new evaluation scheme that can account for phase ambiguities of beat annotation for vocal music.

2- We propose two neural models for the singing beat tracking task. These models leverage pre-trained speech self-supervised models to extract feature embeddings, which are then fed into a linear transformer network to output beat predictions. 

3- We evaluate the proposed models on hundreds of vocal tracks with diverse genres. Experiments show that the proposed models outperform three representative baselines designed or trained for general music beat tracking by a large margin. An ablation study is also performed to investigate the effect of speech self-supervised models over commonly used generic spectral features, and results again show an outperformance by a large margin. 

The remainder of the paper is organized as follows: In Section 2, we describe the proposed vocal beat tracking task with dataset curation and a new evaluation scheme. In Section 3, we describe our proposed models with details. In Section 4, we present experimental comparisons with baselines and an ablation study on the feature extraction front end. Finally, Section 5 concludes the paper.

\section{The Proposed Task}
\label{sec:task}
We propose singing beat tracking as a new MIR task, which aims to design algorithms that can track beats of singing voices in online or offline fashion. In this section, we describe the two cornerstones for this task: datasets and evaluation metrics.

\subsection{Singing Data and Label Generation} \label{subsec:data}
\begin{figure}[htbp]
 \centerline{
 \includegraphics[width=1\columnwidth]{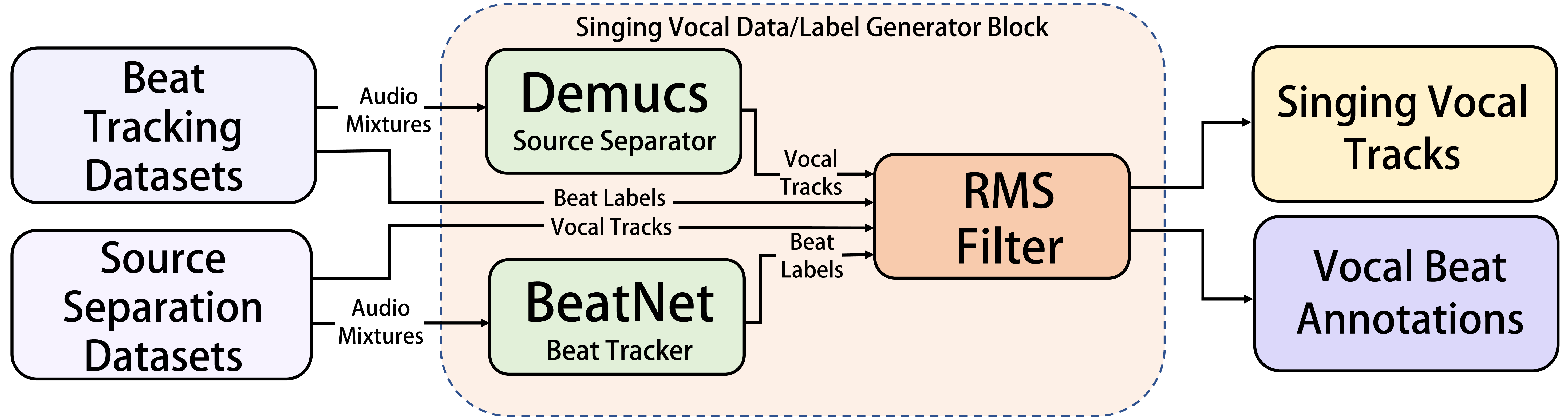}}
 \caption{Singing data and label generation pipeline.}
 \label{preprocess}
\end{figure}

Data for training and evaluating singing beat tracking algorithms require both singing voice recordings and their beat annotations. As there is no existing datasets that meet this requirement, we need to design a mechanism to collect such data. Instead of recording singing voices and annotating their beats manually, we propose to leverage existing MIR datasets and techniques to collect data for singing beat tracking, as illustrated in Figure \ref{preprocess}. 

The first strategy we propose is to use pre-existing music beat tracking datasets in which the beat annotations are available. The idea is to obtain their singing tracks through music source separation. State-of-the-art (SOTA) singing voice separation methods have been shown to achieve outstanding results on popular genres of music, and the separated singing voice contains little interference from the background music. For this task, we employ Demucs Hybrid~\cite{defossez2021hybrid}, one of the superior music source separation models. It uses a waveform-to-waveform convolutional auto-encoder with a U-Net structure and bidirectional LSTM layers, and is designed to separate the music mixture into four sources including bass, drums, vocals, and others. We use the pre-trained model \footnote{https://github.com/facebookresearch/demucs} which was trained on all 150 songs of MusDB~\cite{MUSDB18} dataset. It is noted that source separation is used in some related works e.g.~\cite{desblancs2022zero,chiu2021source,gkiokas2012music} to improve music beat tracking while in this work, we utilize it as one of the suggested systematic ways to generate singing vocal beat tracking datasets. 

The second strategy we propose leverages pre-existing music source separation datasets, in which the isolated vocal tracks are available as a part of those datasets. To obtain beat annotations, we apply BeatNet~\cite{heydari2021beatnet} offline version on the complete songs (i.e., music mixtures). BeatNet is a SOTA online beat tracking method that uses a convolutional recurrent neural network (CRNN) and a hidden Markov model (HMM) decoder to extract music beats and downbeats. We modify the HMM decoder from particle filtering inference to Viterbi algorithm to improve its performance in the offline scenario. The reason that we run BeatNet on music mixtures instead of the separated vocal tracks is because it, the same as all other beat tracking methods, is trained on complete music pieces. 
While BeatNet has been shown to be fairly accurate on many songs with different genres, there are still annotation errors. To fix them, 
we also perform a manual revision by listening to the separated vocal track together with synthesized beeping sounds of the beat annotations and correcting errors.

While the beat annotations are obtained for the entire song, the ground-truth or separated vocal tracks show long chunks of silence due to the inactivity of singing. 
These long silent chunks are not useful and even are distracting for singing rhythmic analysis. Therefore, we normalize the energy of each vocal track using the root-mean-square (RMS), calculate a frame-wise RMS, and set a threshold to detect long silent chunks and split the vocal track into vocal segments. The RMS threshold is set to 0.01, and silent chunks shorter than 8 seconds are kept. 

We apply these two strategies to a total of 8 existing beat tracking and music source separation datasets to obtain a total of 2248 vocal excerpts with beat annotations. The entire length of the vocal segments is 34h 35m. The datasets are summarized in Table 1.

\begin{table}[!b]
  \begin{center}
    \begin{tabular}{L{0.43\columnwidth}C{0.20\columnwidth}C{0.20\columnwidth}}
        \hline
        \textit{Dataset} & \textit{\# Number of vocal excerpts} & \textit{Total Length}  \\
        \hline
        \small Ballroom \cite{Gouyon:2,Krebs:4} $^{*}$ & \small 452 & \small 2 h 38 m   \\
        \hline
        \small GTZAN \cite{Marchand,Tzanetakis} $^{*}$ & \small 741 & \small 5 h 44 m  \\
        \hline
        \small Hainsworth \cite{hainsworth2004particle}$^{*}$ & \small 154 & \small 1 h 47 m \\
        \hline
        \small MUSDB18 \cite{MUSDB18}$^{{\dagger}}$ & \small 263 & \small 6 h 21 m \\
        \hline
        \small Rock Corpus \cite{Clercq}$^{*}$ & \small 315 & \small 9 h 23 m \\
        \hline
        \small RWC pop \cite{goto2002rwc,goto2004development}$^{*}$ & \small 188 & \small 5 h 06 m  \\
        \hline
        \small RWC Royalty free \cite{goto2002rwc,goto2004development}$^{*}$ & \small 29 & \small  19 m \\
        \hline
        \small URSing \cite{li2021audiovisual}$^{{\dagger}}$ & \small 106 & \small 3 h 17 m   \\
        \hline
    \end{tabular}
    \caption{Datasets collected and adapted for singing beat tracking. ${*}$ denotes beat tracking datasets and ${\dagger}$ denotes music separation datasets.}
  \end{center}
\label{datasets}
\end{table}

\subsection{Evaluation Metrics}
\label{sec:task:metrics}
For evaluation metrics, we adopt the commonly used F-measure in beat tracking. A beat is considered correctly detected if it is matched with a ground-truth beat with a time deviation smaller than 70 ms. In addition, we employ three additional metrics including P-score, Cemgil and Goto to provide a more detailed evaluation. Details about these three metrics can be found in~\cite{davies2009evaluation}. 

For many music pieces, vocal tracks can align with the offbeat position (i.e., middle point between two adjacent ground-truth beats) rather that the beats. In fact, it also can be quite natural for humans to clap on the 180-degree phase shifted positions While this inherent ambiguity also exists in some instrumental music, it is much more common for singing voices. Apparently, the off-beat predictions would be evaluated poorly against the ground-truth beat annotations for all of the abovementioned metrics. 


To address this problem,
we propose an additional Phase Inclusive (PI) evaluation scheme with existing metrics. In this scheme, a metric is computed twice, one comparing the predicted beat positions with the ground-truth beat annotations and the other comparing with the 180-degree shifted ground-truth. The maximum is reported as the final metric under the PI evaluation scheme. 

\section{The Proposed Method}\label{sec:method}
\begin{figure}[htbp]
 \centerline{
 \includegraphics[width=1\columnwidth]{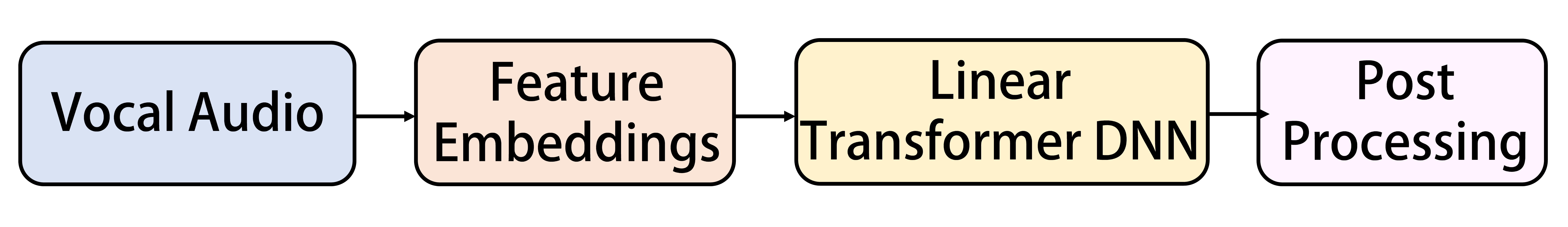}}
 \caption{General pipeline of the proposed method.}
 \label{overview}
\end{figure}

\begin{figure*}[htbp]
 \centerline{
 \includegraphics[width=2.2\columnwidth]{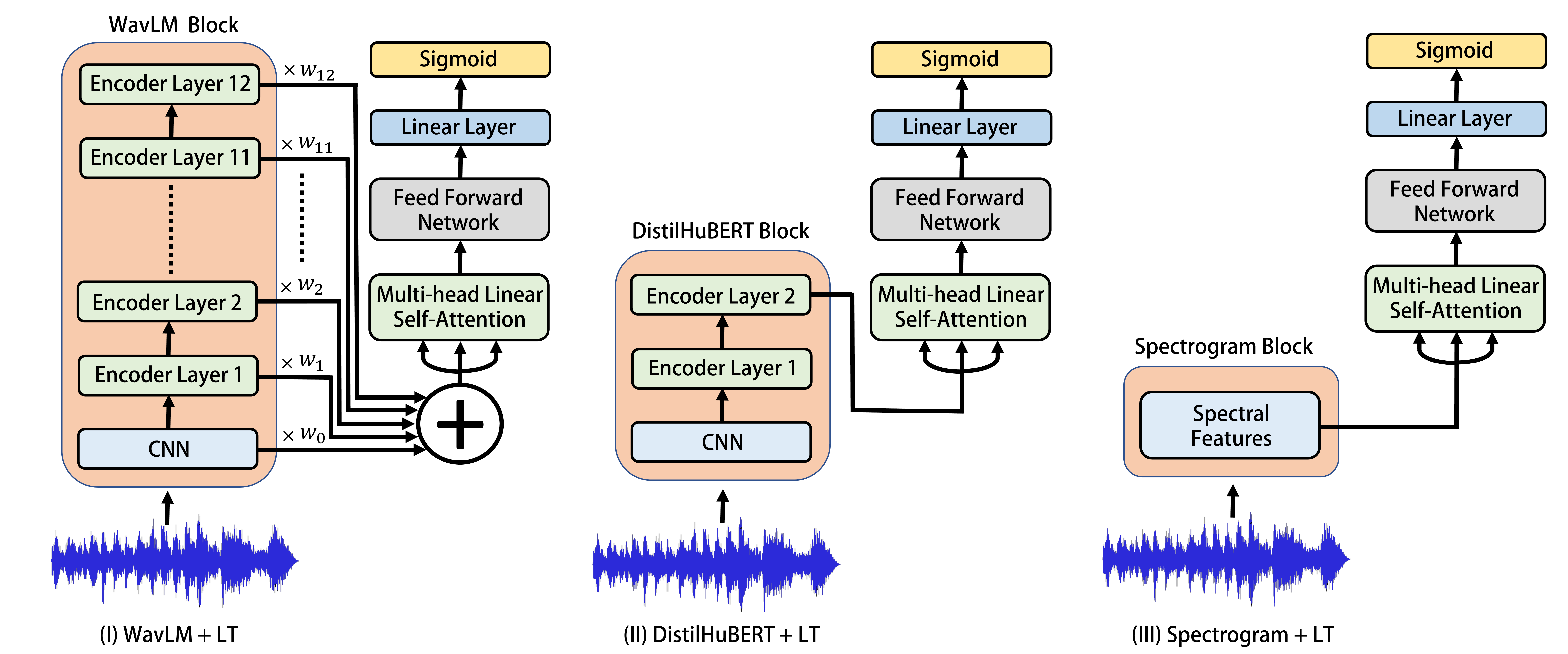}}
 \caption{Neural network structures of the proposed models. (I), (II) and (III) use WavLM, DistilHuBERT and Spectrogram front-ends blocks, respectively, followed by the same linear transformer network.}
 \label{network_structure}
\end{figure*}

To address the singing beat tracking task, we propose two models that take advantage of pre-trained WavLM and DistilHuBERT SSL representations respectively to extract the input features. Then we build the same linear multi-head self-attention network on top of them to fine-tune the models for our task. Finally, a hidden Markov model decoder is used to infer the singing beat positions using the activations provided by the respective neural networks. The source code and system demos are available\footnote{https://github.com/mjhydri/singing-vocal-beat-tracking
} .Figure \ref{overview} demonstrates the overall structure of all proposed models and Figure \ref{network_structure} demonstrates the neural network structures of all proposed models.   

\subsection{Feature Embedding}
\label{subsec:Network architecture}
Neural networks benefit from large quantities of labeled training data. However, in many cases including that of our task, labeled data is much harder to obtain than unlabeled data. Self-supervised learning has emerged as a paradigm to learn general data representations from unlabeled examples and fine-tuning the model on labeled data~\cite{baevski2020wav2vec}. On the other hand, as demonstrated in~\cite{heydari2021beatnet}, selecting appropriate input features plays an important role in music rhythmic analysis tasks. Due to acoustic and linguistic similarities between singing voices and speech, our main assumption is that self-supervised speech representations are helpful as feature embeddings in singing rhythmic analysis. 

SSL achieves great success in speech-related tasks such as Automatic Speech Recognition, Phoneme Recognition and Emotion Recognition. Each task can be addressed by fine-tuning a universal pre-trained self-supervised model or training a neural network on top of the contextualized self-supervised representation of the input. In recent years, several pre-trained self-supervised models (e.g.~\cite{chen2021wavlm,baevski2020wav2vec,hsu2021hubert,peng2022fast,chang2022distilhubert,ling2020decoar}) attempt to provide such universal speech representations. The most successful models such as~\cite{chen2021wavlm,baevski2020wav2vec} usually encode speech audio through a multi-layer convolutional neural network and then mask some chunks of the resulting latent speech representations. Then, a contrastive learning step is performed by feeding the latent representations to a transformer network to distinguish true latents~\cite{baevski2020wav2vec}. 

\textbf{WavLM:}
WavLm~\cite{chen2021wavlm} is one of the recent universal self-supervised pre-trained models on 94k hours of data to solve full-stack downstream speech tasks. It jointly predicts masked speech and performs denoising in the pre-training stage. The denoising module helps to enhance the potential for non-ASR tasks. Moreover, it uses gated relative position bias in its transformer structure to improve capturing sequence ordering of input speech. WavLM has three different versions, Base, Base+, and Large. WavLM Base and WavLM Base+ include 12 encoder layers, 768-dimensional hidden states, and 8 attention heads. The WavLM Large contains 24 encoder layers, 1024-dimensional hidden states, and 12
attention heads. According to the SUPERB ~\cite{yang2021superb}, which is an evaluation benchmark designed to provide a standard and comprehensive testbed for pre-trained models on several speech tasks, WavLM outperforms the other counterparts for several speech downstream tasks.  

Since WavLM model has demonstrated promising results for several downstream speech tasks, we employ it as the front-end model for our first proposed approach to prepare contextualized input embeddings. Among three pre-trained options, we chose the Base+ model, becasue of its better performance than the Base model and similar performance with the Large model on many downstream tasks. 

According to some works on the layer-wise analysis of self-supervised models (e.g.~\cite{prasad2022analyzing,pasad2021layer}), using their last layer's output is not necessarily the best option to leverage the SSL model's maximum capability. This is because the desired feature aspects of the input signal are distributed differently through internal encoder layers. For example,~\cite{pasad2021layer} demonstrated that in Wav2Vec2 self-supervised model, for some properties like word meaning, intermediate layers play a much more important role than the output layer, while for acoustical features, early layers carry the most important information. Therefore, in order to capture different aspects of the input signal, we use a weighted sum of all encoder layers of WavLM in which each weight is a learnable parameter during the training phase. 

\textbf{DistilHuBERT:}
Although WavLM demonstrated the most successful results for several downstream tasks, it is a large model which requires large memory and high pre-training and inference costs imposing a speed bottleneck for the system.  DistilHuBERT~\cite{chang2022distilhubert}, is a novel multi-task
learning framework to distill hidden representations from a HuBERT~\cite{hsu2021hubert} model directly. It is initialized with the HuBERT’s parameters. Then, its prediction heads learn to generate the teacher’s hidden representations by minimizing the loss function. DistilHuBERT reduces the model size by 75\% and increases the inference speed by 73\% from HuBERT while retaining most performance in ten different speech tasks. Moreover, it requires little training time and data, opening the possibilities of pre-training personal and on-device SSL models for speech. 

Therefore, we propose a second model that switches the heavy WavLM front-end model with DistilHuBERT, a much lighter  model, leading to a much faster inference process. DistilHuBERT includes a CNN network and two self-attention layers. According to its evaluation results~\cite{chang2022distilhubert}, DistilHuBERT delivers comparable performance to the teacher HuBERT model with much less computational cost. It is noted that as the training objective of the DistilHuBERT student model is to learn directly from several internal encoder layers of the teacher model, the weighted sum strategy is not needed for this model and its last layer's output is used as the feature embeddings. 



\textbf{Spectrogram:}
We also perform an ablation study to evaluate the effect of using self-supervised models on the system performance by replacing the self-supervised pre-trained feature embeddings with spectral features that are commonly used in SOTA music beat tracking methods~\cite{Bock:1,heydari2021beatnet}. 
These spectral features include log-magnitude mel-frequency spectrogram with three window sizes of 1024, 2048, and 4096 samples and their first-order time differences. 

\subsection{Linear Transformer Network}
\label{subsec:Network architecture}
Transformers achieve remarkable performance in many tasks, but due to their quadratic complexity with respect to the input length, they are prohibitively slow for long sequences~\cite{katharopoulos2020transformers}. To address this limitation and to improve transformers' computational cost, several approaches are proposed. For instance, Linformer~\cite{wang2020linformer} is a model that reduces the transformer complexity in both time and space from $O(N^2)$ to $O(N)$ by approximating the self-attention mechanism by a low-rank matrix. It is noted that $N$ is the sequence length. Another model~\cite{katharopoulos2020transformers}reduces the complexity to linear by expressing the self-attention as a linear dot-product of kernel feature maps and making use of the associativity property of matrix products. It achieves similar performance to vanilla transformers and they are up to 4000x faster on auto-regressive prediction of very long sequences. Another recent model~\cite{lu2021soft} removes the softmax in self-attention by using the Gaussian kernel function to replace the dot-product
similarity without further normalization. It enables a full self-attention matrix to be approximated via low-rank matrix decomposition.

In our proposed method, we build a linear multi-head self-attention layer using~\cite{katharopoulos2020transformers} on top of the feature embedding blocks for all three proposed models.


Our encoder comprises a self-attention layer with four attention heads with query and value dimensions of 192 and a feed-forward network with the size of 1024. The encoder network is followed by a linear layer and a sigmoid layer. The final output can be viewed as the beat salience for the input audio frame. 

To train the proposed models, we use binary cross-entropy with logits between model output and the ground-truth beat annotations as the loss function. Only the linear transformer and its following layers are trained, while the self-supervised front ends are pre-trained and fixed.

\section{Post-Processing}
\label{sec:method:postprocessing}
In order to decode the beat positions, we employed a DBN approximated using HMM from Madmom library~\cite{madmom} with the default parameters. It is noted that the mentioned inference block is shared between all reported models in the evaluation table except BeatRoot~\cite{dixon2007evaluation}.    

\section{Experiments}\label{sec:experiments}

\subsection{Experimental Setup}\label{subsec:Methodology}

We use all of the datasets collected in Section \ref{sec:task} except GTZAN for training the models with 80\% of the songs randomly allocated for training and 20\% for validation. Following several previous music beat tracking works, we keep the entire GTZAN dataset as the test set. It contains 1000 music excerpts covering 10 different music genres, out of which 741 of them contain vocal tracks. It ensures genre diversity in the test set and such a training/test dataset partition reduces the possibility of overfitting as well.Moreover, it is unseen in the training blocks of all compared methods.
However, one drawback of it is that GTZAN's vocal tracks are separated from the whole songs. Hence, they may contain some information leakage from other musical instruments, which may boost the system performance compared to testing on isolated vocal tracks. To address this issue, we took the following steps. First, We used one of the best-performing supervised source separation models~\cite{defossez2021hybrid} and listened through the separated vocal tracks to 
ensure that they do not contain obvious music signal leakage. Second, when tuning the RMS filter, we discarded some sparse, abrupt and short signal segments that are more likely to be percussive sounds instead of vocal signals. Third, we compared with several high performing music beat tracking methods in the evaluation section; The leakage of musical signals into the separated tracks, if any, would improve the performance of these comparison models as well.

For the mentioned reason and given that singing beat tracking is a novel MIR task with no specifically design existing models, in this paper, we compare our proposed models against general music beat tracking models including two supervised models, BeatNet\cite{heydari2021beatnet} and Böck~\cite{Bock:1} and an unsupervised signal-processing model, BeatRoot~\cite{dixon2007evaluation}. BeatNet is the SOTA online joint beat, downbeat, and meter tracking model that uses convolutional recurrent neural networks and particle filtering. It can also operate in an offline fashion by switching its particle filtering inference block with an offline hidden Markov model decoder. Böck uses a recurrent neural network to predict beat and downbeat saliences of each audio frame, and uses a Dynamic Bayesian Network (DBN) to infer beat and downbeat positions from the salience function. In our implementation, we use the same HMM decoder as that in BeatNet to replace the DBN. BeatRoot is a unsupervised method with a multiple agent architecture that simultaneously considers several different hypotheses of beat positions and tempi.

\subsection{Training Details}\label{subsec:Training Details}


To train the proposed models, we freeze the self-supervised front-end blocks and train the rest of the model including the multi-head self-attention layer and its following feed-forward and linear layers. For the WavLM+LT model, in addition to the abovementioned items, the 12 weights to combine different encoder layer's output are also trained.

The mentioned weights are initialized as ones and the rest of the weights and biases are initialized randomly. 
We use the Adam optimizer with a learning rate of $5 \times 10^{-5}$ and a batch size of 10. The batches comprise 15-second long excerpts randomly sampled from all audio files in the training set. As the audio files have different length, to ensure a fair distribution, we sample with a probability that is proportional to the length of the files.

\subsection{Results and Discussions}
\label{subsec:Results and Discussion}

To assess the computational cost of the models, we report the inference speed for all of them. The reported numbers are the average computational time for all 741 excerpts with the average duration of ~28 seconds each. Note that we measure the speed of all methods using CPU processing on the same Windows machine with an AMD Ryzen 9 3900X CPU and 3.80 GHz clock. Table \ref{evaluation} demonstrates the evaluation results.

\begin{table*}[]
\centering
\begin{tabular}{ccccccc}
                                                                                  & Method            & F-Measure      & P-Score        & Cemgil         & Goto           & Comp. Time \\ \hline
\multirow{3}{*}{Baseline}                                                         & BeatNet           & 0.243          & 0.327          & 0.173          & 0.003          & 0.13 (s)          \\
                                                                                  & BeatRoot          & 0.301          & 0.394          & 0.22           & 0.066          & \textbf{0.03 (s)} \\
                                                                                  & Böck              & 0.171          & 0.195          & 0.122          & 0.009          & 1.56 (s)          \\ \hline
\multirow{3}{*}{Proposed}                                                         & WavLM + LT        & \textbf{0.733} & \textbf{0.704} & \textbf{0.618} & \textbf{0.560} & 4.09 (s)          \\
                                                                                  &  DistilHuBERT + LT & 0.703          & 0.668          & 0.593          & 0.516          & 1.83 (s)          \\
                                                                                  &  Spectrogram + LT  & 0.454          & 0.438          & 0.367          & 0.223          & 0.32 (s)          \\ \hline
\multirow{3}{*}{\begin{tabular}[c]{@{}c@{}}Proposed \\ (PI Results)\end{tabular}} &  WavLM + LT        & 0.745          & 0.715          & 0.627          & 0.574          & 4.09 (s)          \\
                                                                                  &  DistilHuBERT + LT & 0.721          & 0.684          & 0.608          & 0.537          & 1.83 (s)          \\
                                                                                  &  Spectrogram + LT  & 0.489          & 0.477          & 0.391          & 0.265          & 0.32 (s)
                                                                                  \\ \hline
\end{tabular}
    \caption{Average performance and speed across segments of several methods on the GTZAN separated vocal tracks.}
\label{evaluation}
\end{table*}

Table \ref{evaluation} shows the evaluation results of the baselines and the proposed models using the four metrics presented in Section \ref{sec:task:metrics}. It also includes the Phase Inclusive (PI) evaluation scheme for the proposed models. Several interesting observations can be made. 
First, all of the three proposed models outperform the three baselines by a large margin. This difference is especially pronounced for more strict metrics such as Goto~\cite{davies2009evaluation}. This confirms our hypothesis that vocal tracks show very different patterns from complete songs, and the baseline models that are trained on complete songs do not perform as well as the proposed models that are trained on the vocal tracks. This hypothesis is further validated by the fact that BeatRoot outperforms BeatNet and Böck, while the latter two models are shown to outperform BeatRoot in previous studies on complete songs~\cite{Bock:1,heydari2021beatnet}. 
It is noted that BeatNet and Böck are data-driven approaches and are trained on complete songs, while BeatRoot is a signal processing model without training. The mismatch between complete songs and vocal tracks does impose a strong negative bias to the data-driven baselines.

Second, the two proposed models that leverage speech SSL feature embeddings (i.e., WavLM+LT and DistilHuBERT+LT) improves over Spectrogram+LT by a large margin, and this improvement is even bigger than that from the best baseline to Spectrogram+LT. This shows the significant advantage of using feature embeddings learned on speech data and suggests that this advantage is even more significant than training on the vocal tracks. As the spectrogram features used in Spectrogram+LT are the same as those in BeatNet and Böck and transformers have been shown to outperform RNNs in various tasks, it is reasonable to believe that even if BeatNet and Böck are trained on the vocal tracks, their performance would be only comparable to that of Spectrogram+LT. 

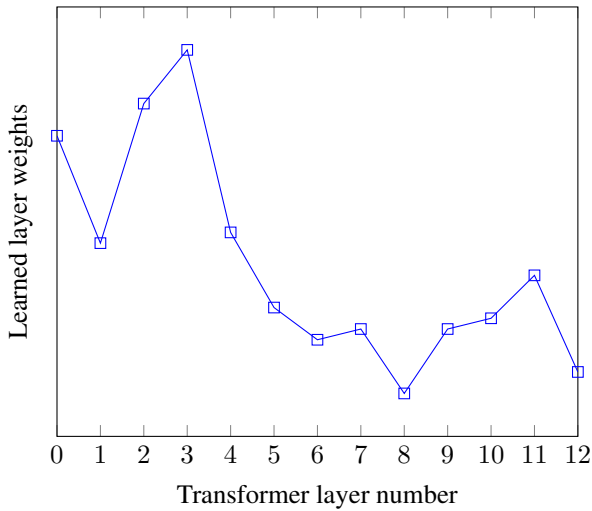
\begin{figure}
\centering
\begin{tikzpicture}
\begin{axis}[
    xlabel={Transformer layer number},
    ylabel={Learned layer weights},
    ylabel style= {at={(+0.11,0.5)}},
    xmin=0, xmax=12,
    ymin=0.73, ymax=0.77,
    xtick={0,1,2,3,4,5,6,7,8,9,10,11,12},
    ytick={0,0.2,0.4,0.6,0.8,1},
    ymajorgrids=true,
    grid style=dashed,
]
\addplot[
    color=blue,
    mark=square,
    ]
    coordinates {
    (0,0.758)(1,0.748)(2,0.761)(3,0.766)(4,0.749)(5,0.742)(6,0.739)(7,0.740)
    (8,0.734)(9,0.740)(10,0.741)(11,0.745)(12,0.736)
    };
    \end{axis}
    \end{tikzpicture}
\caption{Illustration of the learned weights for different encoder layers of WavLM in the WavLM+LT model.}
\label{layerwise}
\end{figure}

Third, WavLM+LT delivers the best scores for all of the evaluation metrics except the computation time, showing that stronger feature embeddings lead to better vocal beat tracking accuracy. However, 
the attention layers' complexity in WavLM is quadratic to the length of input; while the average computational time (4.09 seconds) is acceptable for vocal segments (~28 s long on average) in this experiments, as the length increases, the computational time will soon become prohibitive.
Figure \ref{layerwise} shows the learned weights of different encoder layers of the WavLM SSL model. It can be seen that earlier layers generally contribute more to forming the feature embeddings compared to latter layers. Layers 3, 2 and 1 contribute the most. Note that Layer 0 represents the CNN network output.   

Fourth, with slight performance drop, the second proposed model which employs LT and DistilHubert delivers the next best performance among all models. DistilHubert uses a much lighter structure with fewer parameters than WavLM, making it ideal for time-sensitive inferences.

Finally, the PI evaluation scheme shows a slight improvement of all of the proposed models on all accuracy metrics. In particular, the improvement of Spectrogram+LT is greater than that of the other models. This suggests that the SSL pretrained speech embedding features are helpful to reduce the phase shift ambiguity.

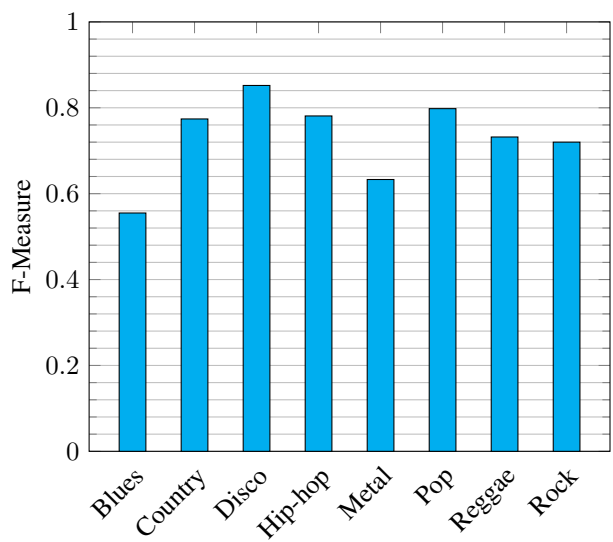
\begin{figure}[!b]
\centering
\begin{tikzpicture}
        \hspace*{-0.3cm}
        \begin{axis}[
            symbolic x coords={Blues, Country, Disco, Hip-hop, Metal, Pop, Reggae, Rock},
            xtick=data,
            ylabel=F-Measure,
            xticklabel style={rotate=45,anchor=north east},
            ylabel style= {at={(+0.05,0.5)}},
            ymajorgrids,yminorgrids,minor y tick num=4,
            ymin=0,ymax=1
          ]
            \addplot[ybar,fill=cyan] coordinates {
                (Blues, 0.555) 
                (Country, 0.774) 
                (Disco, 0.852 ) 
                (Hip-hop, 0.781 )
                (Metal, 0.633 )
                (Pop, 0.798 )
                (Reggae, 0.732 )
                (Rock, 0.720 )
            };
        \end{axis}
         \label{fig3}
    \end{tikzpicture}
\caption{F-measure performance of WavLM + LT model on the GTZAN separated vocal tracks for different genres.}
\label{evaluation-genre}
\end{figure}

Figure \ref{evaluation-genre} shows the performance of our proposed WavLM + LT model for different music genres. According to the table, the best and worst performances belong to \emph{disco} and \emph{blues} respectively. One reason for that may be the stronger and punchier stresses in \emph{disco} vocal tracks comparing to that of \emph{blues} tracks. Another important reason may be the different syncopation ratio among vocal track of different music genres. Syncopation is a musical term that defines the placement of rhythmic stresses or accents where they would not normally occur, making part or all of a tune or piece of music off-beat~\cite{hoffman2009syncopation}. Therefore, a higher syncopation ratio leads to a more difficult beat tracking process. Nate that in Figure \ref{evaluation-genre} we only report the evaluation results of the genres that contain at least 80 separate singing tracks. It leaves out the \emph{classical} and \emph{jazz} genres which have only two and zero separated tracks.


\section{CONCLUSION}\label{sec:Conclusion}
In this paper we introduced singing beat tracking as a novel MIR task. We proposed two approaches to obtain labeled data for the mentioned task and introduced two methods to accomplish the mentioned task using pre-trained speech self-supervised models and multi-head linear self-attention networks. We also conducted an ablation study to investigate the effect of speech self-supervised models on the system performance. Experiments on a collection of vocal tracks with diverse genres show that the proposed models that combine self-supervised models and transformers outperform three representative baselines designed or trained for beat tracking for general music. The ablation study also shows that the self-supervised speech embeddings outperform generic spectral features that are commonly used in music beat tracking. 
In the future we will extend this research to cover real-time applications and more challenging singing tracks such as human humming and operas.   

\section{ACKNOWLEDGEMENT}\label{sec:Acknowledgement}
This work was supported by National Science Foundation grant No. 1846184.

\bibliography{ISMIRtemplate}

%
%
%
%
%

\end{document}